# Ultrafast THz probe of photo-induced polarons in lead-halide perovskite

*Eugenio Cinquanta[1,2,\*], Daniele Meggiolaro[3,4], Silvia G. Motti[5], Marina Gandini[5], Marcelo J. P. Alcocer[1,6], Quinten Adriaan Akkerman[7,8], Caterina Vozzi[2], Liberato Manna[7], Filippo De Angelis[3,4], Annamaria Petrozza[5\*] and Salvatore Stagira[1,2]*

1) Dipartimento di Fisica, Politecnico di Milano, Milano, Italy

2) Istituto di Fotonica e Nanotecnologie, Consiglio Nazionale delle Ricerche, Milano, Italy

3) Computational Laboratory for Hybrid/Organic Photovoltaics, CNR-IMS, Perugia, Italy

4) D3-CompuNet, Istituto Italiano di Tecnologia, Genova, Italy

5) Center for Nano Science and Technology, Istituto Italiano di Tecnologia, Milano, Italy;

6) Solid State Physics and NanoLund, Lund University, P.O. Box 118, SE-221 00 Lund, Sweden

7) Department of Nanochemistry, Istituto Italiano di Tecnologia, Via Morego 30, 16163 Genova, Italy

8) Dipartimento di Chimica e Chimica Industriale, Università degli Studi di Genova, Via Dodecaneso, 31, 16146, Genova, Italy

**Corresponding Author**

*Eugenio Cinquanta, eugenioluigi.cinquanta@polimi.it

*Annamaria Petrozza, annamaria.petrozza@iit.it

**ABSTRACT** We study the nature of photo-excited charge carriers in $CsPbBr_3$ nanocrystal thin films by ultrafast optical pump - THz probe spectroscopy. We observe a deviation from a pure Drude dispersion of the THz dielectric response that is ascribed to the polaronic nature of carriers; a transient blueshift of observed phonon frequencies is indicative of the coupling between photogenerated charges and stretching-bending modes of the deformed inorganic sublattice, as confirmed by DFT calculations.

Hybrid Organic-Inorganic Perovskites (HOIPs) are "soft-lattice" ionic semiconductors with a direct band gap that presents very interesting physical properties including superconductivity, magnetoresistance, ionic conductivity, ferroelectricity and piezoelectricity, which are of great importance for microelectronics and telecommunication [1]. Recently, metal halide perovskites have been the subject of intensive studies, since they have been identified as the base materials for high efficient solar cells [2,3].

Despite this active research, there are aspects of the physics of these materials that are still not completely understood. For instance, HOIPs behave as defect-free semiconductors with all the scattering processes involving charge carriers (defects, phonons, charges) mitigated by the screened Coulomb potential, leading to long carrier lifetimes, long diffusion length and slow electron-hole recombination [4]. Unexpectedly, HOIPs show only a moderate carrier mobility that looks conflicting with the previous characteristics. The presence of large polarons – resulting from the dielectric electron-phonon coupling combined with the light effective masses for bare carriers – has been proposed as a possible explanation for the limited carrier mobility in this class of materials [4]. However, no direct experimental fingerprint of the large polaron in the transport properties of HOPIs have been reported to date.

In this Letter, by combining Ultrafast THz spectroscopy with Density Functional Theory (DFT) calculations, we demonstrate the presence of large polarons in the dielectric response of thin films of $CsPbBr_3$ (a prototypical lead-halide perovskite nanocrystals). We observe a distinctive coupling between the photo-generated charges and dipole-active bending modes of the deformed $PbBr_3$ lattice in the transient THz response. The deviation from a free-carriers dispersion of the transient optical conductivity and the influence of the photoinjected charge on the lattice modes relaxation reveal the polaronic nature of the carriers.

Static and ultrafast THz spectroscopy have been fruitfully adopted for identifying the HOIPs low-frequency optical phonons, charge-carrier recombination rates and hot electron/hole transfer [5-10]. Electronic excitations in most polar semiconductors are strongly coupled to the lattice vibrations. This coupling can lead to carrier trapping or it can induce a certain degree of localization that may result in the formation of polarons, thus changing the nature of the primary generated carrier and affecting transport and recombination mechanisms. Phonon interactions can also strongly perturb the Coulomb correlations between carriers through the modulation of the effective dielectric permittivity, eventually leading to phonon "dressing" of the resultant excitonic states. HOIPs are characterized by a small charge carrier effective mass that, in a first approximation, would lead to Bloch states. The softness of the lattice however results in a dielectric function that departs significantly from a pure Drude-like behavior [5,6].

In this framework, the presence of large polaron has been mentioned by C. Wehrenfennig *et al.* to explain dipole oscillations in the pump-induced THz conductivity spectra [11]. In addition, time-resolved optical Kerr spectroscopy and femtosecond impulsive stimulated Raman spectroscopy allowed to indirectly address the presence of large polarons in HOIPs with different constituents [12,13].

Here we performed optical-pump THz-probe spectroscopy exploiting our terahertz time-domain spectrometer driven by 25-fs pulses from a 1-kHz Ti:sapphire laser with a center wavelength of 790 nm (1.6 eV). THz pulses with a 0.1-2.7 THz bandwidth were generated and detected by electro-optic sampling in two 1-mm thick ZnTe crystals. For exciting the sample, 400-nm (3.1 eV) laser pulses focused to a spot size diameter of 3 mm at a fluences of 85 $\mu J/cm^2$ with a corresponding density of absorbed photons of $5.8 \cdot 10^{17}$ *photons*/$cm^3$ were used. Our set up can detect a transient dynamic induced in the sample up to a pump-probe delay of about 200 ps.

The investigation has been performed on CsPbBr$_3$ nanocrystal thin films. The nanocrystals have been synthesized as reported in [3] and have an average size of about 30 nm, well beyond the critical size for observing quantum confinement effects. The THz probe pulse has an electric field peak amplitude of tens of kV·m$^{-1}$ and about 1-ps-long optical cycle; thus, it can displace charge carriers on length scales much shorter than the average nanocrystal size. We can hence reasonably consider our outcomes as bulk properties of single crystals. Further technical details concerning the thin films preparation are reported in the Supplementary Information [14].

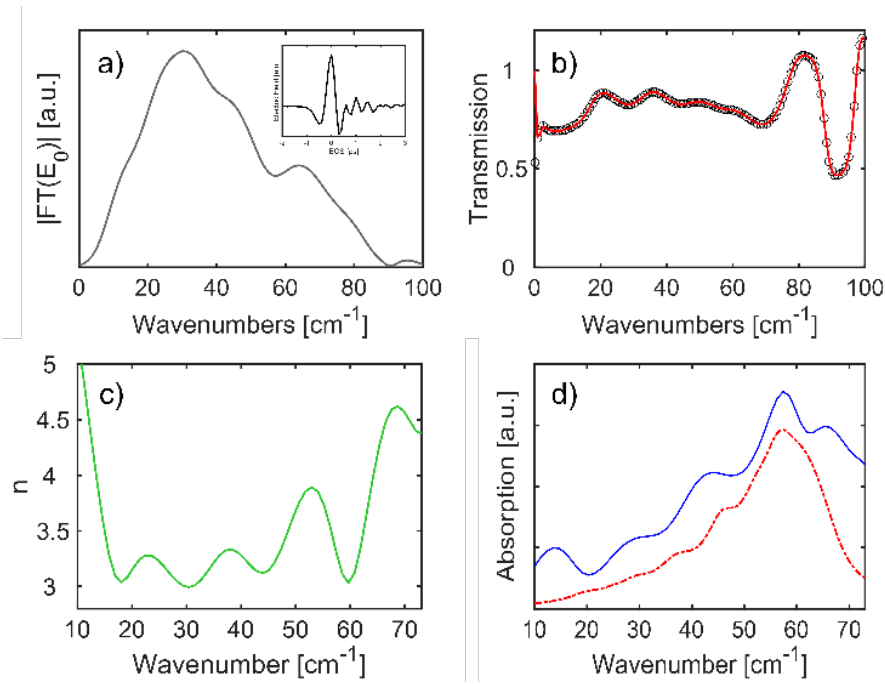

**Figure 1.** a) THz probe spectrum used in the optical pump/THz probe experiment with its waveform in the time-domain reported in the inset; b) Experimental transmission of the unexcited sample $T(\omega) = \frac{T_{Sample}(\omega)}{T_{Reference}(\omega)}$ (black circle) and corresponding fit with the model transmission (continuous red curve); c) retrieved real part of the complex refractive index of the sample; d) calculated (red dashed curve) and experimental (blue continuous curve) THz absorption spectrum of the sample.

We firstly investigated the *static properties* of the perovskite thin films to elucidate the presence of phonons that are expected to be involved in large polaron formation. Figure 1a shows the THz probe spectrum and the corresponding measured electric field (inset). Figure 1b reports the experimental transmission of the unexcited sample $\frac{T_{Sample}(\omega)}{T_{Reference}(\omega)}$ (black circle). In order to extract the static dielectric properties of the thin film, this transmission has been fitted with a model transmission (continuous red line) obtained by considering the transmission of the THz electric field through the system composed by the thin film and the substrate [15]. Details concerning the THz data extraction are reported in the Supplementary Information [14].

Figure 1c and 1d show the real part of the refractive index and the absorption spectrum of a 3-μm thick perovskite film in the 10-75 cm$^{-1}$ spectral range (continuous green and blue curves respectively) acquired at room temperature and extracted from the raw data through the above-mentioned procedure. The refractive index and the absorption spectrum show the presence of four phonon features at 27, 42, 57 and 63 cm$^{-1}$. Phonon resonances show typical Lorentzian line shapes: the absorption spectrum has a peak in correspondence of each phonon energy, whereas the refractive index shows an inflection point at the same energy. These results agree with previous studies on lead-halide perovskites with different constituents [5,6]. These peaks can be finely retrieved in the DFT-calculated THz absorption spectrum (dashed red curve) and are related to the superimposition of several IR-active pure bending and mixed stretching-bending modes of the Pb-Br cage [12]. Note that the calculated absorption spectrum shows a good qualitative agreement with the experimental one; however, anharmonicity, local polar fluctuation, dynamic disorder and cation rotational unlocking are not included in the model. At room temperature, all these contributions can partly affect the phonon frequencies with inhomogeneous peak broadening, resulting in the observed discrepancy between the experimental data and the model [6,16,17].

The large polaron is expected to appear because of coupling of the photoinjected charge with soft lattice phonons [12,18]. Previous ultrafast THz spectroscopy studies of strongly-confined ~10 nm large colloidal CsPbBr$_3$ nanocrystals performed in the 0.5-5 THz frequency range reported a Drude response, although small oscillations appeared in the transient optical conductivity spectra [9]. A further study on the same system showed how the transient THz permittivity is characterized by the presence of Lorentzian line shapes that are attributed to the presence of strong carrier-phonon coupling and to a multi-phonon process responsible for the hot carrier relaxation [10]. The role of electron-phonon coupling on the pump-induced conductivity has also been studied by THz spectroscopy in MAPbI$_3$ thin films [7,19].

Remarkably, in our measurements we can clearly identify the large polaron fingerprint in the transient dielectric response of CsPbBr$_3$ nanocrystal thin films. Figure 2a reports the transient optical conductivity acquired at 3 and 100 ps after the photoexcitation (blue circle and black squares, respectively) in the spectral range from 15 to 65 cm$^{-1}$. Data analysis has been performed with the same retrieval procedure adopted for the static dielectric properties. We observe the presence of three peaks in the real part of the transient optical conductivity spectra. These peaks correspond to inflection points in the imaginary part of the transient optical conductivity (see SI) and this response clearly indicates the presence of pump-induced dipole oscillations characterized by Lorentzian line shapes. The red curve in Figure 2a is the fitting of the experimental data with the Drude-Lorentz model $\sigma_\text{f}(\omega) = \frac{nq^2}{m_\text{p}} \cdot \frac{1}{\gamma - \iota\omega} + \sum_m \frac{g_m \omega}{\iota(\omega_m^2 - \omega^2) + \Gamma_m \omega}$, where $n$ is the density of carriers, $q$ is the elementary charge, $m_p$ is the carrier mass, $\gamma$ is the scattering rate, $g_m$ is the oscillator strength, $\Gamma_m$ is the FWHM of the Lorentzian curve, $\omega_m$ and $\omega$ are the THz angular frequencies of the $m$-phonon and of the probe electric field, respectively. This model hence describes the contribution to the optical conductivity of the polaronic carriers. We adopted $m_p = 3.5 \cdot m^*$ for the carrier mass ($m^*$=0.21·$m_e$ is the bare carrier effective mass and $m_e$ is the rest electron mass) in the Drude term to account for the larger mass of the polaronic carriers [12,20]. The resulting mobility $\mu = \frac{q \cdot \tau_s}{m_p}$ of ~60 cm$^2$/V·s (scattering time $\tau_s$=1/$\gamma \sim$ 20 fs as extracted from the fit) is in good agreement with the experimental values reported in the literature [20] and with the predicted upper bound for the polaron mobility of 38 cm$^2$/V·s calculated with the Feynman-Osaka polaron model [12].

In order to elucidate the nature of the vibrational modes associated to the photoinduced lattice deformations, we analyzed our results with the aid of DFT calculations performed on the orthorhombic phase of CsPbBr$_3$. Details of the calculation are reported in the Supplementary Information [14]. The overall lattice deformation induced by a positive charge within the unit cell was calculated using the hybrid PBE0 functional [21,22] and fixing cell parameters. The response of the system to the addition of electrons has not been considered here, due to the negligible lattice deformations calculated both at the GGA and hybrid level of theory [12]. The difference of the Cartesian positions of the ions between the neutral and positive energy minima configurations is the variable representing the lattice distortion. Upon injection of a hole, an average shortening of the Pb-Br bond lengths in the octahedra is expected (-0.03 Å) coupled to an increase of the Br-Pb-Br angle of ~10° [12,23]. This relaxation stabilizes the lattice by 0.14 eV and pushes the system toward a more 'cubic' symmetry, with cations following such a distortion and occupying their corresponding cubic positions in the cell. Based on these observations, the response of the lattice to the positive charge is mainly related to the activation of Br-Pb-Br stretching and bending modes.

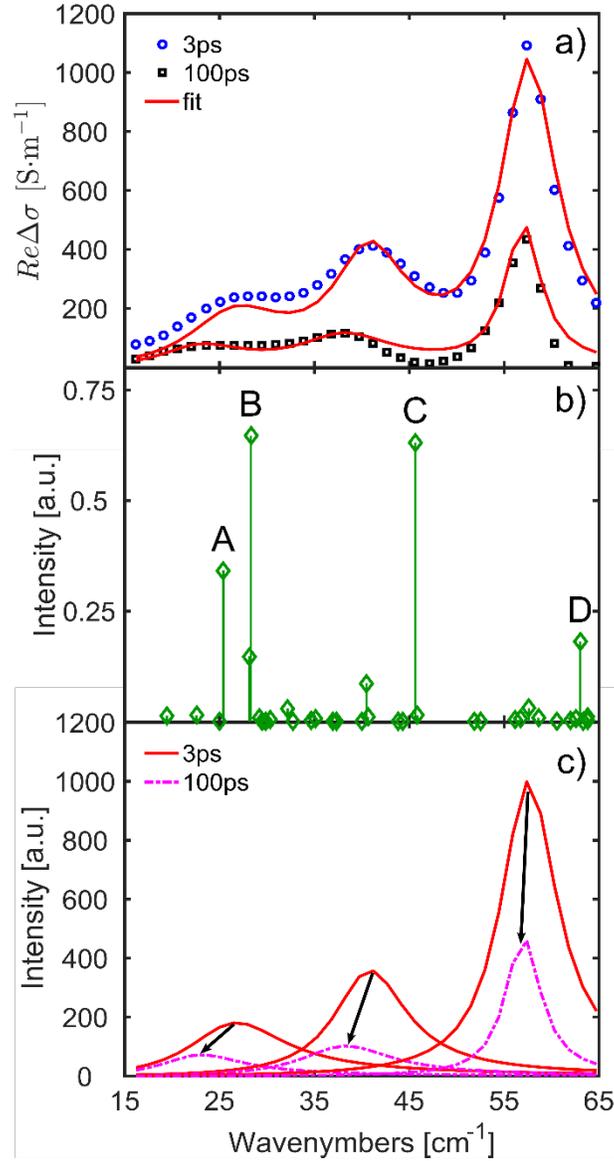

**Figure 2.** a) Frequency-resolved transient optical conductivity at a pump fluence of 85 μJ/cm$^2$ (5.6·10$^{17}$ photons/cm$^3$). Symbols represent the real part of the transient optical conductivity at a pump-probe delay of 3 ps (blue dots) and 100 ps (black squares), the red lines are a fitting of the experimental data with the Drude-Lorentz model; b) calculated coefficients of the electron-phonon coupling related to far-IR phonons involved in the large polaron formation; A, B, C and D are the labels of the most intense phonons mode coupled to the photo-generated carriers; c) real part of the Lorentzian fit at 3 and 100 ps (continuous red and dashed magenta lines, respectively).

Figure 2b shows the calculated normalized coefficients of the displacement vectors projected onto the normal modes of the crystal in the relevant spectral region. The most intense displacement activity is predicted for four Pb-Br-Pb pure bending modes at 25, 28, 46 and 63 cm$^{-1}$, labelled as A, B C and D, that are strongly coupled with the injected charge. These values are in very good agreement with the three phonons at 27, 42 and 58 cm$^{-1}$ extracted from the Drude-Lorentz fit of the experimental spectra at 3 ps after the photoexcitation. Indeed, according to the theoretical prediction, these phonon modes are responsible for the formation of the large polaron in the photoexcited lattice. In this respect, we want to highlight that the presence of pump-induced phonon resonances in the transient THz response is not common for non-polar/polar semiconductors, III-V compounds and van der Waals materials. Typical bulk non-polar semiconductors, like Silicon, present free carriers (Drude-like) transient optical conductivity [24]. Bulk GaAs, a polar-semiconductor, shows a strong coherent coupling of the LO phonon at ~8 THz with the plasmon through Coulomb interactions and a pure Drude response at lower energies (0.5 – 2.5 THz), while GaAs nanowires show only surface plasmon resonance [25-27]. Van der Waals semiconductors, like MoS$_2$, show free carriers together with the presence of charged excitons (Trions) [28]. On the other hand, the lattice softness of lead halide perovskites, 10 times softer with respect to Si and GaAs, is the crucial ingredient for the appearance of the coupling between the photo injected charge and phonon modes [29].

To gain further insights into the polaronic fingerprint, we analyze the spectra acquired at different pump-probe delays. Figure 2c shows the real part of the Lorentzian fits at 3 and 100 ps (continuous red and dashed magenta curves, respectively). At a pump-probe delay of 100 ps, the intensity is decreased, and the phonons are slightly red shifted (24, 40 and 57 cm$^{-1}$) as compared to the 3 ps delay. This red shift can be explained considering that photogenerated holes in CsPbBr$_3$ nanocrystals distort the PbBr lattice more than electrons [12,23]. Indeed, once the photoexcitation creates a hole in the top of the valence band – which has an anti-bonding character –the Pb-Br bond becomes stiffer and the lattice shrinks, resulting in a blueshift of phonon frequencies. As the photo injected charge density starts to decrease, the lattice can expand again, and the Pb-Br bond softens. The observed peak frequencies at 3 ps and 100 ps in Figure 2d are thus consistent with a lattice compression in the first ps after the photoexcitation and the subsequent expansion during the carrier recombination.

If the photo injected carriers were not coupled to the lattice by the polaron formation, their relaxation would not impact lattice mode frequencies. Given the redshift observed as a function of the pump-probe delay and the agreement between the DFT model and the deconvolution of pump induced conductivity spectra, we claim that the observed features are the fingerprint of presence the large polaron in lead-halide perovskites.

To further explore the carrier dynamics, we studied the frequency-averaged dielectric response of the sample for different optical pump intensities, corresponding to different carrier densities; this response is obtained by recording the transient change of the THz field peak as a function of the pump-probe delay. Figure 3a reports the transient electric field $-\frac{\Delta E(t_p,0)}{E_0}$ at pump fluences of 0.3, 1.1 and 2.2·$10^{18}$ photons/cm$^3$ respectively, where $\Delta E(t_p,0)$ is the change in the transmitted THz field detected at different pump-probe delays $t_p$ in correspondence of the peak of the THz waveform located at $t_{eos}= 0$ ($t_{eos}$ is the electro-optical sampling delay) and $E_0$ is the maximum of the THz field transmitted by the unexcited sample. With our experimental setup we are not able to follow the monomolecular decay processes that typically occurs on the *ns* time scale for this class of materials [3]. Within 200 ps the sample shows an initial fast decay of the transient electric field change with a time constant of about ~15 ps and a slower one of hundreds of ps (Figure 3a). The slow component falls well within the band-to-band carrier recombination regime as probed by photoluminescence measurements and often reported in the literature [3]. The fast decay component becomes dominant when the initial carrier density $N_0$ is larger than $0.5 \cdot 10^{18}/cm^3$. The relationship between $N_0$ and the absorbed photon density $N_{ph}$ is calculated as $N_0 = \varphi \cdot 2 \cdot N_{ph}$, where the photon to carrier ratio was set to $\varphi = 1$. At these densities, trap states are saturated (see Supporting Information) and many-body processes start to play a role, thus this fast dynamic can be assigned to the increase of Auger-like recombination.

Figure 3b shows the build-up of the $\frac{\Delta E(t_p, 0)}{E_0}$ signal up to 3 ps at 0.3 and $1.1 \cdot 10^{18}$ photons/cm$^3$. The observed formation time is in good agreement with recent outcomes obtained on different perovskite systems. In [30], S. A. Bretschneider *et al* modeled the measured build-up dynamics for different perovskites thin films with a simplified model that considers the initial hot electron cooling and the subsequent large polaron formation. The cooling of the hot electron distribution occurs on a sub picosecond time scale and is temperature and excess-energy dependent ($\tau_{cool}$). The large polaron formation, on the other hand, occurs at 400 fs and is temperature and excess-energy *independent* ($\tau_{pol}$). We hence adopt this model and fit the build-up dynamics with a 4 level rate equations system, were the top-level A is populated at time 0. Level A and B are blind state to take into account cooling and polaron formation; the whole dynamics was fitted using level C population. The formation time $\tau_{pol}$ was fixed at 400 fs according to [30] and the dynamics were convoluted with a Gaussian of 100 fs FWHM (pump pulse duration). The resulting fits are reported in Figure 3b as continuous red curves and nicely describe the experimental data. Within this model we obtain for the cooling time 450 and 540 fs at 0.3 and $1.1 \cdot 10^{18}$ photons/cm$^3$ respectively, that are in good agreement with the values ($\tau_{cool} \sim 470$ fs) reported for CsPbI$_3$ samples at room temperature and photoexcited by 400 nm pump [30]. The build-up dynamics measured for the photoconductivity further demonstrate the presence of large polarons in photoexcited perovskites thin films.

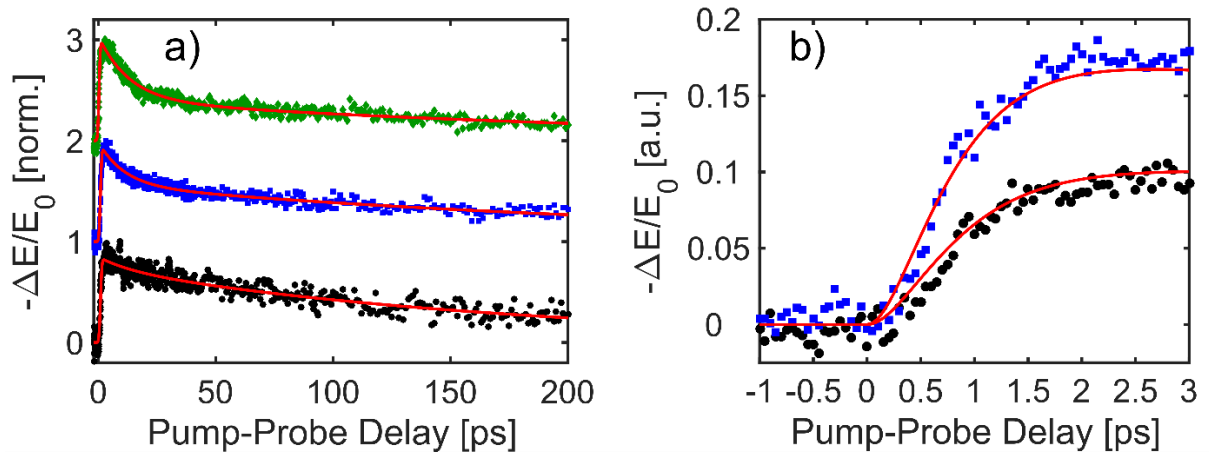

**Figure 3.** a) Transient change of the THz electric field peak transmitted by the CsPbBr$_3$ sample at 0.3, 1.1 and 2.2·10$^{18}$ photons/cm$^3$ carrier densities together with the biexponential fit (red continuous line); b) Transient dynamics in the first 3 ps after the photoexcitation at 0.3 and 0.5·10$^{18}$/cm$^3$ (black dots and blue squares respectively); the red lines are the fit of the build-up dynamics to the rate equation

In summary, we directly unveiled the presence of large polarons in CsPbBr$_3$ nanocrystal thin films by optical pump-THz probe spectroscopy combined with state-of-the-art DFT calculations. The pump-induced dielectric response of the perovskite thin films is decorated by Lorentzian line shapes that imply the presence of dipole-active vibrational modes coupled to the photo injected charges due to electron-phonon interaction. Our finding represents a direct experimental evidence of polaronic transport in lead-halide perovskites thin films and can contribute to the understanding of the exceptional physics of these compounds that makes them a rich playground for the implementation of next generation optoelectronics device.

**AUTHOR INFORMATION**

*E-mail: eugenioluigi.cinquanta@polimi.it

*E-mail: annamaria.petrozza@iit.it13

**ACKNOWLEDGMENT**

E.C. thanks D. Viola for the help for the fit of the THz build-up dynamics. E. C., M. J., P. A. and S. S. acknowledge financial support from Fondazione CARIPLO, Project "GREENS" (ref. no. 2013-0656). E.C. and C. V. acknowledge funding from the European Research Council Starting Research Grant UDYNI (Grant No. 307964) from the Italian Ministry of Research and Education (ELI project - ESFRI Roadmap), and from Regione Lombardia through the project FEMTOTERA (ID: CONCERT2014-008).